\documentclass{article}

\usepackage{arxiv}

 \usepackage[hyphens]{url}
\PassOptionsToPackage{hyphens}{url}
\usepackage{breakurl}

\usepackage{breakcites}
\usepackage{graphicx}
\usepackage{pgfplots}
\usepackage{placeins} 

\usepackage[utf8]{inputenc} 
\usepackage[T1]{fontenc}    
\usepackage{hyperref}       
\usepackage{url}            
\usepackage{booktabs}       
\usepackage{amsfonts}       
\usepackage{nicefrac}       
\usepackage{microtype}      
\usepackage{lipsum}		
\usepackage{graphicx}

\usepackage[nolist,nohyperlinks]{acronym}
\usepackage{hyperref}
\usepackage{cleveref}

\makeatletter
\newcommand*{\org@overidelabel}{}
\let\org@overridelabel\@verridelabel
\@ifpackagelater{acronym}{2015/03/21}{
  \renewcommand*{\@verridelabel}[1]{%
    \@bsphack
    \protected@write\@auxout{}{\string\AC@undonewlabel{#1@cref}}%
    \org@overridelabel{#1}%
    \@esphack
  }%
}{
  \renewcommand*{\@verridelabel}[1]{%
    \@bsphack
    \protected@write\@auxout{}{\string\undonewlabel{#1@cref}}%
    \org@overridelabel{#1}%
    \@esphack
  }%
}
\makeatother
\Crefformat{figure}{#2Fig.~#1#3}
\Crefmultiformat{figure}{Fig.~#2#1#3}{ and~#2#1#3}{, #2#1#3}{ and~#2#1#3}
\Crefformat{table}{#2Table~#1#3}
\Crefmultiformat{table}{Tabels~#2#1#3}{ and~#2#1#3}{, #2#1#3}{ and~#2#1#3}
\Crefformat{equation}{#2Equation~#1#3}
\Crefmultiformat{equation}{Equations~#2#1#3}{ and~#2#1#3}{, #2#1#3}{ and~#2#1#3}
\crefname{lstlisting}{listing}{listings}
\Crefname{lstlisting}{Listing}{Listings}

\usepackage{tikz}
\usepackage{tikzpeople} 
\usetikzlibrary{positioning}
\usetikzlibrary{arrows,arrows.meta}
\usetikzlibrary{arrows,positioning,automata,shadows,fit,shapes,calc}
\usetikzlibrary{arrows.meta,calc,decorations.markings,math,arrows.meta,patterns,backgrounds}

\newcommand*\circledg[1]{\tikz[baseline=(char.base)]{%
            \node[shape=circle,fill=black!40!white,inner sep=0.1pt] (char) {\footnotesize \textcolor{white}{#1}};}}

\usepackage{xcolor}

\definecolor{lime}{HTML}{A6CE39}
\DeclareRobustCommand{\orcidicon}{
	\begin{tikzpicture}
	\draw[lime, fill=lime] (0,0)
	circle [radius=0.16]
	node[white] {{\fontfamily{qag}\selectfont \tiny ID}};
	\draw[white, fill=white] (-0.0625,0.095)
	circle [radius=0.007];
	\end{tikzpicture}
	\hspace{-2mm}
}

\foreach \x in {A, ..., Z}{%
	\expandafter\xdef\csname orcid\x\endcsname{\noexpand\href{https://orcid.org/\csname orcidauthor\x\endcsname}{\noexpand\orcidicon}}
}

\usepackage[nolist,nohyperlinks]{acronym}
\begin{acronym}
  \acro{ACK}{acknowledgement}
  \acro{ARP}{Address Resolution Protocol}
  \acro{CPS}{Cyber Physical System}
  \acro{CPU}{Central Processing Unit}
  \acro{CSA}{Component Security Assurance}
  \acro{DoS}{Denial of Service}
  \acro{DDoS}{Distributed Denial of Service}
  \acro{DuT}{Device under Test}
  \acro{EDSA}{Embedded Device Security Assurance}
  \acro{FIFO}{First In -- First Out}
  \acro{GPIO}{General Purpose Input Output}
  \acro{GPOS}{General Purpose Operating System}
  \acro{ICS}{Industrial Control System}
  \acro{ICMP}{Internet Control Message Protocol}
  \acro{MCU}{Microcontroller Unit}
  \acro{OS}{Operating System}
  \acro{PLC}{Programmable Logic Controller}
  \acro{RT}{Real-Time}
  \acro{RTOS}{Real Time Operating System}
  \acro{RTU}{Remote Telemetry Unit}
  \acro{SCADA}{Supervisory Control and Data Acquisition}
  \acro{SYN}{synchronize}
  \acro{TCP}{Transmission Control Protocol}
  \acro{UDP}{User Datagram Protocol}
  \acro{IP}{Internet Protocol}
  \acro{GUI}{Graphical User Interface}
\end{acronym}

\title{Analysis of Industrial Device Architectures \\for Real-Time Operations \\under Denial of Service Attacks}

\author{ \href{https://orcid.org/0000-0002-4532-731X}{\includegraphics[scale=0.06]{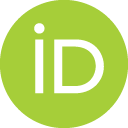}\hspace{1mm}Florian Fischer}\\
	Hochschule Augsburg, Germany\\
	\url{www.hsainnos.de}\\
	\texttt{florian.fischer@hs-augsburg.de} \\
	\And
  \href{https://orcid.org/0000-0003-4550-7422}{\includegraphics[scale=0.06]{orcid.png}\hspace{1mm}Matthias Niedermaier}\\
  	Hochschule Augsburg, Germany\\
  	\url{www.hsainnos.de}\\
  	\texttt{matthias.niedermaier@hs-augsburg.de} \\
    \And
    Thomas Hanka\\
    	Hochschule Augsburg, Germany\\
    	\url{www.hsainnos.de}\\
    	\texttt{thomas.hanka@hs-augsburg.de} \\
      \And
    Peter Knauer\\
    	Hochschule Augsburg, Germany\\
    	\url{www.hsainnos.de}\\
    	\texttt{peter.knauer@hs-augsburg.de} \\
      \And
    Dominik Merli\\
    	Hochschule Augsburg, Germany\\
    	\url{www.hsainnos.de}\\
    	\texttt{dominik.merli@hs-augsburg.de} \\
}

\hypersetup{
pdftitle={Analysis of Industrial Device Architectures for Real-Time Operations under Denial of Service Attacks},
pdfsubject={Research paper},
pdfauthor={Florian Fischer, Matthias Niedermaier, Thomas Hanka, Peter Knauer, Dominik Merli},
pdfkeywords={Industrial Control Systems, Real-Time, Denial of Service, Network-based Attack, Flooding},
}

\newcommand\copyrighttext{%
  \footnotesize \textcopyright Analysis of Industrial Device Architectures for Real-Time Operations under Denial of Service Attacks -- First published in the 22nd International Conference on Information and Communications Security (ICICS 2020)
}
\newcommand\copyrightnotice{%
\begin{tikzpicture}[remember picture,overlay]
\node[anchor=south,yshift=10pt] at (current page.south) {\fbox{\parbox{\dimexpr\textwidth-\fboxsep-\fboxrule\relax}{\copyrighttext}}};
\end{tikzpicture}%
}

\begin{document}
\maketitle

\begin{abstract}
  More and more industrial devices are connected to \acs{IP}-based networks, as this is essential for the success of Industry 4.0.
  However, this interconnection also results in an increased attack surface for various network-based attacks.
  One of the easiest attacks to carry out are \acs{DoS} attacks, in which the attacked target is overloaded due to high network traffic and corresponding \acs{CPU} load.
  Therefore, the attacked device can no longer provide its regular services.
  This is especially critical for devices, which perform \ac{RT} operations in industrial processes.
  To protect against \acs{DoS} attacks, there is the possibility of throttling network traffic at the perimeter, e.g. by a firewall, to develop robust device architectures.
  In this paper, we analyze various concepts for secure device architectures and compare them with regard to their robustness against \acs{DoS} attacks.
  Here, special attention is paid to how the control process of an industrial controller behaves during the attack.
  For this purpose, we compare different schedulers on single-core and dual-core Linux-based systems, as well as a heterogeneous multi-core architecture under various network loads and additional system stress.
\end{abstract}


\acresetall

\copyrightnotice

\section{Introduction}
Modern industrial devices, like \acp{PLC}, are more and more connected to \acs{IP}-based network structures due to the trend of Industry 4.0, which is based on network-enabled machines, actuators and sensors.
In addition to controlling a \ac{CPS}, the connectivity enables features, like easy configuration, remote data collection, web services, updating of firmware or uploading of control programs.
These features increase the productivity and comfort of use, but the wide accessibility also enlarges the attack surface of the industrial controllers through various network-based attack vectors \cite{rubio2019}.
Furthermore, the historical network isolation by air gaps no longer applies, making components a possible target for network-based attacks.

The third industrial revolution was founded on the use of \acp{PLC}, first introduced by Modicon 1969, which automate the manufacturing process by digital programming and provided a huge gain of productivity \cite{drath_horch_2014}.
These devices were designed without considering any security aspects for their connectivity, since the devices were separated from IT network infrastructures by the concept of air gap.
This means a physical segregation of the \ac{ICS} network from other networks.
As a consequence, the air gap for the industrial processes is nowadays not always guaranteed and other mechanisms are required, since the communication between IT and OT networks arises due to Industry 4.0.

Many processes in the industrial context require a control loop, that must react fast and within a certain time, e.g. the fill up process in a bottling plant.
The \acp{PLC}, which control the processes, are working in a cyclic manner, i.e. they repeatedly execute a control program e.g. every 1\,ms.
Deviation of this cyclic execution of the process control program can lead to bad consequences, like too much or too little fill quantitiy.
These systems must provide a deterministic behavior, i.e. a known latency jitter from stimulus to response within the industrial control program.
This deterministic control can be provided by \ac{RT} capable devices.

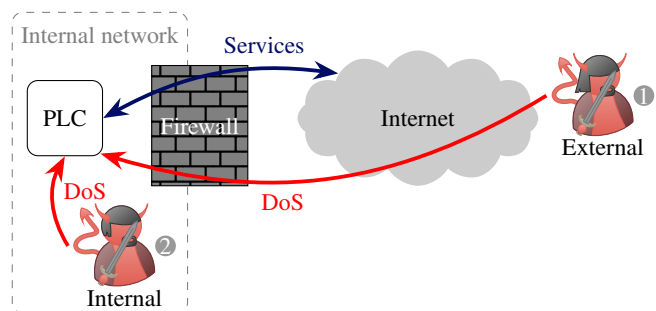
\begin{figure}[htb]
  \centering
  \begin{tikzpicture}[node distance=0.1cm,
    auto,
    block/.style={
      rectangle,
      draw=black,
      align=center,
      rounded corners
    }
  ]
  \coordinate (a) at (0,2.6);    
  \coordinate (b) at (7.66,2.9); 
  \coordinate (c) at (1.3,0.9);	 
  \coordinate (d) at (5.2,2.6);	 
  \coordinate (e) at (2.3,2.5);	 

\definecolor{darkblue}{rgb}{0,0.05,0.4}
  \node[block, draw, dashed, color=gray, align=center, minimum width=2.0cm, minimum height=4.0cm, anchor=south west, text depth = 3.4cm] at (-0.2,0) () {\footnotesize Internal network};

  \node[alice,shirt=red!30!black,female,sword,devil,minimum size=0.8cm, text width=1.8cm, align=center](ext) at (b) {{\footnotesize External}};
  \node[shape=circle,fill=black!40!white, inner sep=0.1pt, anchor=west] at (ext.east)() {\footnotesize \textcolor{white}{1}};
  \node[bob,shirt=red!30!black,sword,devil,minimum size=0.8cm, text width=1.8cm, align=center](int) at (c) {{\footnotesize Internal}};
  \node[shape=circle,fill=black!40!white, inner sep=0.1pt, anchor=west] at (int.east)() {\footnotesize \textcolor{white}{2}};

  \node[block, draw, align=center, minimum width=1cm, minimum height=1cm, anchor=west] at (a) (plc) {\footnotesize  \acs{PLC}};
  \node[rectangle, draw, pattern=bricks, preaction={fill=gray}, minimum width=1cm, minimum height=1.6cm] at (e) (firewall) {\footnotesize \textcolor{white}{Firewall}};
  \node [cloud, cloud puffs=12,cloud puff arc=120, aspect=2, inner ysep=0.1cm,minimum width=3.2cm, minimum height=1.8cm, fill=black!20!white] at (d) (internet) {\footnotesize  Internet};

  \draw [-{Stealth[scale=1.0]}, bend angle=-24, line width=0.5mm, color=red, bend right]  ([xshift=-10pt]ext.west) to node[below,xshift=-18pt,yshift=-2pt]{\footnotesize \acs{DoS}} ([xshift=-1pt,yshift=1pt]plc.south east);
  \draw [{Stealth[scale=1.0]}-{Stealth[scale=1.0]}, bend angle=18,line width=0.5mm,color=darkblue, bend left]  (plc.east) to node[above,xshift=18pt,yshift=4pt]{\footnotesize Services} (internet.north west);
  \draw [-{Stealth[scale=1.0]}, bend angle=30, line width=0.5mm, color=red, bend left]  ([xshift=-10pt]int.west) to node[right,yshift=4pt]{\footnotesize \acs{DoS}} (plc.south);

  \end{tikzpicture}
  \caption{\acs{DoS} attack scenarios on \acp{PLC} addressed in this work.}
  \label{fig:attackscenario}
\end{figure}

Recent research shows, that the cyclic operations of several commercial control devices can be influenced by network-based attacks.
Niedermaier et al. presented, that high network traffic loads can affect the cyclic execution of common \ac{PLC} devices up to complete system failures \cite{niedermaier18woot}.
These kind of flooding attacks enable even less experienced attackers with access to the industrial network to influence the cyclic execution of control programs and can thereby cause disturbance to controlled physical processes.
Two different types of attackers are considered, as shown in \Cref{fig:attackscenario}.
For instance an external attacker~\circledg{1} can perform a \ac{DoS} attack on a service, that is accessible from external networks, like the Internet,
The second attacker type is an internal attacker~\circledg{2} who, in addition to an intended attack with flooding, could also trigger an \ac{DoS} attack unintentionally e.g. by executing a network scan.
This attacker type differentiation assumes proper firewall configuration, so no flooding attack or scanning, e.g. for asset management, is possible from external networks.

Vendors of industrial components have to make decisions about the underlying architecture design in an early development state.
This early decision must already consider proper protection against \ac{DoS} attacks.
The recent industrial security standard IEC 62443-4-2 requires \ac{DoS} Protection from secure  industrial devices.
Therefore a Security-by-Design approach for modern industrial control devices is obligatory, to achieve robust regular execution of a control task, when network connectivity is necessary.
A secure architecture for industrial control devices is required, which is also robust against various kinds of network flooding attacks and under certain system loads.
To evaluate suitable system architectures for future \ac{PLC} designs, influences of certain stress factors on \ac{RT} execution must be considered.

The \ac{GPOS} Linux is a common choice for industrial components, due to need of implementing demanding \acp{GUI} or the possibility, to use open source libraries.
In addition Linux can be used for commercial products and supports multiple hardware architectures.
The implementation of features is commonly less costly on fully featured operating systems, like Linux, and therefore development duration and costs are far less, than on systems, based on bare metal or \ac{RTOS} \cite{reghenzani_massari_fornaciari_2019}.
Linux is a \ac{GPOS}, but by applying the \texttt{preempt\_rt} patch, it becomes \ac{RT} capabilities \cite{oliveira_oliveira_2015}.
This combination addresses the requirement for a fully fledged \ac{OS} and the real-time needs for demanding industrial control processes.

The contribution of this paper is the systematic measurement of influences on a control program under network load and CPU stress scenarios, executed on different device architectures.
Further, we discuss the measurement results and the consequences of architectural choices on the robustness of industrial components against \ac{DoS} attacks.

The paper is organized as follows.
First we provide a summary of related work in \Cref{sec:relatedwork}, followed by the presentation of technical background about \ac{RT} systems, \acp{PLC} and network-based attacks in \Cref{sec:technicalbackground}.
We present the selection of architectures in \Cref{sec:architecture}.
\Cref{sec:methodology} describes the method of our measurements and the measurement setup.
Further, the test cases and the results are shown in \Cref{sec:results}, followed by a discussion in \Cref{sec:discussion}.
A final conclusion is provided in \Cref{sec:conclusion}.

\section{Related Work}
\label{sec:relatedwork}

Influences of \ac{DoS} attacks on \ac{ICS} components are still a topic in research and were investigated in previous works.
Long et al. analyzed \ac{DoS} attacks on network-based control systems and how this degrades their performance almost two decades ago \cite{long_wu_hung_2005}.
Although this work discusses \ac{DoS} attacks on \acp{PLC}, the results are based on simulation data and no investigations of real devices is done.
This theoretic work handles effects of delay on communication between \ac{PLC} and remote devices and not the effects in connection with system load, produced by receiving many packets.

Markovic et al. also provide measurements of performance degradation under \ac{DDoS} attacks, based on simulation data \cite{markovic-petrovic_stojanovic_2013}.
In comparison to these works, we want to provide realistic measurement data by investigating influences on electrical controls of real devices.

Niedermaier et al. presented measurement of common of the shelf \acp{PLC} under certain network loads \cite{niedermaier18woot}.
The measurements aim for physical influences of network-based flooding attacks, but lacks architecture comparison to provide a robust network-enabled \ac{PLC} design.
We use similar measurement routines, but also discuss architecture approaches and their suitability for future robust \acp{PLC}.

Recent research already discusses robust hardware architecture for \ac{ICS}.
Niedermaier et al. presents a dual controller setup, which separates the control task and the communication part by hardware \cite{niedermaier_merli_sigl_2019}.
This architecture design requires implementation of a custom dual controller setup and lacks the capability to run a full fledged \ac{OS}, like Linux.
This concept is not comparable to architectures, which provide the features of a full \ac{OS} for functionality next to the process control, like in the architecture designs within our work.

Lelli et al. discuss the deadline scheduler within Linux and compare the percentage of missed deadlines with \texttt{SCHED\_FIFO} and \texttt{SCHED\_OTHER} under certain system load scenarios \cite{lelli_scordino_abeni_faggioli_2015}.
They come to the conclusion, that \texttt{SCHED\_DEADLINE} is suitable for hard \ac{RT} tasks, if the taskset can be partitioned and the per-core load is \textless 1.
However no results for full system utilization and other sources of load, e.g. network traffic, are discussed within their work.
This worst case scenario for \ac{RT} execution is analyzed by our measurements.

On Linux-based \ac{RT} systems the measurement tool cyclictest is a common method to determine the kernel latency on all cores of a \ac{DuT} \cite{cyclictest}.
Linutronix runs continuous tests on multiple hardware platforms with this tool and applies a defined load during the tests \cite{linutronix}.
The test routines contains certain network communication load, but lack investigation of influences during network-based attacks, like flooding.

In further work there is already investigation of jitter on RT patched Linux systems, measured on digital outputs\cite{ArthurSiro}\cite{Brown}\cite{MossigeMorten}, but none covers external influences, like high network loads to these systems.

In summary, related work does not cover demanding cyclic execution under heavy network-based attacks.
Therefore we focused our work on creating a test methodology, which covers these scenarios.

\section{Technical Background}
\label{sec:technicalbackground}
This section discusses the topics \ac{RT}, \acp{PLC} and network-based attacks.

\subsection{Real Time}

To control a \ac{CPS}, control devices need to provide \ac{RT} operations.
\ac{RT} by far does not mean fast processing, but deterministic and in-time execution of a certain task.
A specific process has to complete and provide its results within a time mark, known as deadline.
\ac{RT} systems can be classified in the categories soft, firm and hard \cite{kopetz_2011}.
This classification is done by the consequences accompanied with the miss of a \ac{RT} system deadline.
Deadline misses on soft \ac{RT} systems lead to a degradation of the event value after the deadline.
Misses on a firm \ac{RT} system degrade the value of the event to zero after the deadline.
A single deadline miss on a hard \ac{RT} system can lead to a complete fault of the controlled process.
This can cause catastrophic consequences, like out-of-control production processes, destruction of production environment and gear or even hazard to human beings.

\subsection{Programmable Logic Controllers (PLC)}

\begin{figure}[htb]
  \centering
  \begin{tikzpicture}[node distance=0.1cm,
    auto,
    block/.style={
      rectangle,
      draw=black,
      align=center,
      rounded corners,
      dashed
    }
  ]
  \coordinate (a) at (0,0);    
  \coordinate (b) at (4,0);    
  \coordinate (c) at (4,-2.6);	  
  \coordinate (d) at (0,-2.6);	  
  \coordinate (e) at (2,-1.3); 

  \node[circle,draw,dashed,inner sep=0pt,minimum size=2.1cm, text width=1.9cm,align=center] (1)
        at (a) {1. Read inputs};
  \node[shape=circle,fill=black!40!white, inner sep=0.1pt, anchor=south] at ([yshift=0.2cm]1.south)() {\footnotesize \textcolor{white}{1}};
  \node[circle,draw,dashed,inner sep=0pt,minimum size=2.1cm, text width=1.9cm,align=center] (2)
        at (b) {2. Program execution};
  \node[shape=circle,fill=black!40!white, inner sep=0.1pt, anchor=south] at ([yshift=0.2cm]2.south)() {\footnotesize \textcolor{white}{2}};
  \node[circle,draw,dashed,inner sep=0pt,minimum size=2.1cm, text width=1.9cm,align=center] (3)
        at (c) {\small 3. House-\\keeping, e.g. networking};
  \node[shape=circle,fill=black!40!white, inner sep=0.1pt, anchor=south] at ([yshift=0.2cm]3.south)() {\footnotesize \textcolor{white}{3}};
  \node[circle,draw,dashed,inner sep=0pt,minimum size=2.1cm, text width=1.9cm,align=center] (4)
        at (d) {4. Write outputs};
  \node[shape=circle,fill=black!40!white, inner sep=0.1pt, anchor=south] at ([yshift=0.2cm]4.south)() {\footnotesize \textcolor{white}{4}};
  \node[circle,inner sep=0pt,minimum size=2.3cm, text width=2.0cm,align=center] (5)
        at (e) {Cycle time};

  \draw [-{Stealth[scale=1.0]}, bend angle=20, bend left]  (1) to node[above]{} (2);
  \draw [-{Stealth[scale=1.0]}, bend angle=20, bend left]  (2) to node[above]{} (3);
  \draw [-{Stealth[scale=1.0]}, bend angle=20, bend left]  (3) to node[above]{} (4);
  \draw [-{Stealth[scale=1.0]}, bend angle=20, bend left]  (4) to node[above]{} (1);

  \draw [-{Stealth[scale=1.0]}, bend angle=50, bend left, shorten >=0.2cm,shorten <=0.2cm, line width=1mm, color=gray] ([xshift=-1.1cm]e) to node[above]{} ([xshift=1.1cm]e);
  \draw [-{Stealth[scale=1.0]}, bend angle=50, bend left, shorten >=0.2cm,shorten <=0.2cm, line width=1mm, color=gray]  ([xshift=1.1cm]e) to node[above]{} ([xshift=-1.1cm]e);
  \end{tikzpicture}
  \caption{Program execution on a cycle orientated \acs{PLC}.}
  \label{fig:cycle}
\end{figure}
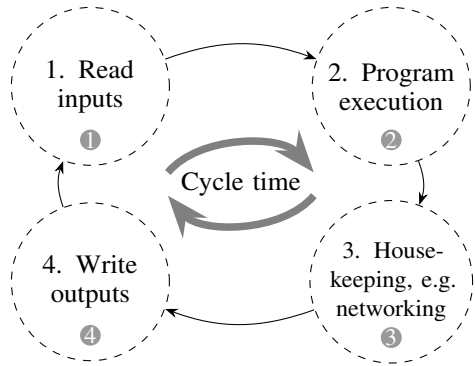

\acp{PLC} nowadays are the main devices to control \ac{ICS} and therefore have to provide \ac{RT} capability.
The conditions, that must be met are highly depending on the physical process, controlled by the device.
Therefore control devices must be prepared, to provide the required \ac{RT} capability even in worst case scenarios, like high system load or during a network-based attack.
A control program is executed on \acp{PLC} in a cyclic manner and processes the four steps illustrated in \Cref{fig:cycle}.

The \emph{read inputs} stage \circledg{1}~ handles the read of digital or analog inputs on the \ac{PLC}.
The \emph{program execution} stage \circledg{2}~ handles the execution of the cyclic control program.
The \emph{housekeeping} part \circledg{3}~ can service communication requests, internal checks or diagnostic functionality.
The \emph{write outputs} stage \circledg{4}~ handles the writing of the logic values back to the electric analog or digital output pins of the \ac{PLC}.
This control program is executed periodically within the configured cycle time.
Deviation of this cycle time can cause a delayed response on stimuli from the \ac{CPS} and therefore disturb the \ac{RT} controlling.

\subsection{Network-based Attacks}

A known method to achieve deviation on \ac{RT} devices is network flooding.
Network packets produce certain system load at the receiving device on arrival, even if no payload is handled, since information of the underlying protocols from some message types are still processed.
Parsing and interpreting this information already consumes CPU time, even if the receiver has no listening service running and is not awaiting certain packets.
Flooding massive amount of packets containing \ac{TCP}, \ac{SYN} packets or \ac{ARP} requests for instance are producing high amount of system load and aim for \ac{DoS} on the device.
This externaly triggered system load can be used to disturb the regular execution of the receiving device.
Filtering and blocking of certain network packets e.g. with firewall rules can mitigate these influences, but degrades also the throughput of the network traffic, e.g. by limiting the SYN packet rate.

\section{Architectures}
\label{sec:architecture}
Robust and distortion free execution of the cyclic industrial control program is an essential feature of \acp{PLC}.
For this reason, this work analyzes and compares the influences of flooding attacks on different device architectures.

The implementation of \ac{RT} capable devices can be based on various design approaches.
The focus in this paper is on full open source implementations and therefore the considered solutions lack proprietary concepts and software.
In the following the architecture concepts, which are used as the \acp{DuT} within the measurement procedure, are introduced.

Two architectures, which execute the critical task on a \texttt{preempt\_rt} patched Linux are defined:
(S) a single-core system and (D) a dual-core system.
We configure these architectures to use the real time schedulers \texttt{SCHED\_FIFO} (First-In-First-Out), \texttt{SCHED\_RR} (Round Robin) and \texttt{SCHED\_DEADLINE}.
In addition (D) also pins the critical process to CPU2, while the network kernel process runs on CPU1.

The third variant (C) is a special architecture design, based on a dual-core Linux system and a co-processor.
This co-processor handles the critical control task, while network communication is handled on the Linux system.

These architecture concepts result in the following test cases:
\begin{itemize}
  \item [\textbf{(S)}] Single-core system running Linux
    \begin{itemize}
      \item [\textbf{(SF)}] using \texttt{SCHED\_FIFO}
      \item [\textbf{(SR)}] using \texttt{SCHED\_RR}
      \item [\textbf{(SD)}] using \texttt{SCHED\_DEADLINE}
    \end{itemize}
  \item [\textbf{(D)}] Dual-core system running Linux
  \begin{itemize}
    \item [\textbf{(DF)}] using \texttt{SCHED\_FIFO} and pinned execution on CPU2
    \item [\textbf{(DR)}] using \texttt{SCHED\_RR} and pinned execution on CPU2
    \item [\textbf{(DD)}] using \texttt{SCHED\_DEADLINE} and pinned execution on CPU2
  \end{itemize}
  \item [\textbf{(C)}] Dual-core system running Linux with additional co-processor
\end{itemize}

For single-core (S) and dual-core (D) test cases a Raspberry Pi 4 is used.
The Linux image is created with the buildroot environment \cite{buildroot} and enabled \texttt{preempt\_rt} patch on Kernel Version 4.19.113.
Since the background processes have influence on process latencies, this minimal setup is used.
On the Linux-based \ac{DuT}, the static priority schedulers \texttt{SCHED\_FIFO} and \texttt{SCHED\_RR} are configured with the highest possible priority of 99.
The \texttt{SCHED\_DEADLINE} configures its three parameters the following.
Runtime is set to 100\,000\,ns, which is the execution time assigned to the task within a period, while its deadline and period are set to 1\,000\,000\,ns.

For the co-processor test case (C), a development board with the STM32MP1 \ac{MCU} from ST Microelectronics is used \cite{stm32mp1}.
The \ac{MCU} consists of a dual-core Cortex A7 CPU with additional Cortex M4 co-processor.
This device provides a test case for hardware separated execution.

\section{Methodology and Measurement Parameters}
\label{sec:methodology}
The architectures selected for this work are used to implement a minimal \ac{RT} industrial process.
Thereby, we want to provide measurements and comparison of influences on the regular execution of a task on these architectures during network-based attacks and additional synthetic system loads.
The results can give advice for future designs of robust \acp{PLC}.

\begin{figure}[htb]
  \centering
	\begin{tikzpicture}
	\tikzstyle{every node}=[font=\small]
	\begin{axis}[
	width=8cm,
	height=3.0cm,
	x axis line style={-stealth},
	y axis line style={-stealth},
	ymax = 4,xmax=8.5,
	axis lines*=center,
	ytick={0,3.3},
	xlabel={Time in ms $\rightarrow$},
	ylabel={Output},
	xlabel near ticks,
	ylabel near ticks,
	legend style={at={(1.3,0.75)},anchor=north,legend columns=1}]
	\addplot+[thick,mark=none,const plot,color=blue]
	    coordinates
	    {(0,0) (0,3.3) (1,0) (2,3.3) (3,0) (4,3.3) (5,0)};
	\addplot+[thick,mark=none,const plot,dashed,color=blue]
	    coordinates
	    {(5,0) (6,3.3) (7,0) (8,3.3)};
	\addplot+[thick,mark=none,const plot,dashed,color=red]
	    coordinates
	    {(5.5,0) (6.6,3.3) (8.2,0)};
	\legend{Regular,Expected,Delayed}
	\end{axis}
	\end{tikzpicture}
  \caption{Impacts on cycle time and output signal.}
  \label{fig:delay}
\end{figure}
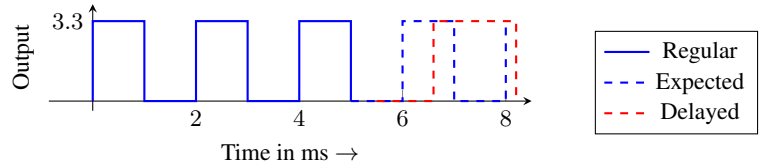

To represent a common cyclic execution of industrial control programs, a elementary periodical program, that inverts the logic level of a physical output within every cycle, is used.
So the stages \emph{program execution} \circledg{2} and \emph{write outputs} \circledg{4} are present.
Other execution stages, like \emph{reading inputs} \circledg{1} or \emph{housekeeping} \circledg{3} are not necessary and thus omitted within our analysis.
The minimal control program is provided for the Linux-based systems (S) and (D), as well as on the bare metal system (C).
The cycle time within our control program is set to 1\,ms in all configured test cases, as this is a common minimal cycle time for off-the-shelf \acp{PLC}.
This results in an digital output signal, as illustrated in \Cref{fig:delay}.

The regular signal is a square wave signal, which changes the output state every 1\,ms.
If no influence occurs, the square wave signal is continuing as expected.
In contrast to this, the square wave signal delays or keeps the current output value, when network flooding influences the control program.

\subsection{Setup}
\label{subsec:setup}
\begin{figure}[htb]
  \centering
  \begin{tikzpicture}[node distance=1cm,
    auto,
    block/.style={
      rectangle,
      draw=black,
      align=center,
      rounded corners,
      dashed
    }
  ]
  \coordinate (a) at (0,2.1);
  \coordinate (b) at (5,2.1);
  \coordinate (c) at (0,0);

  \node[block, draw, align=center, minimum width=2.0cm, minimum height=1.5cm, anchor=west] at (a) (dut) {\acs{DuT}~~~~~\\~};
  \node[rectangle, draw, align=center, minimum width=0.8cm, minimum height=0.5cm, anchor=south, color=gray] at (dut.south) (duteth) {\footnotesize ETH};
  \node[rectangle, draw, align=center, minimum width=0.6cm, minimum height=0.5cm, anchor=east, color=gray] at ([yshift=+0.2cm]dut.east) (dutout) {\footnotesize out};

  \node[block, draw, align=center, minimum width=2.2cm, minimum height=1.5cm, text width=2.0cm, anchor=west] at (b) (capture) {\footnotesize ~~~~~~Logic\\ ~~~~~~analyzer \\~};
  \node[rectangle, draw, align=center, minimum width=0.8cm, minimum height=0.5cm, anchor=south, color=gray] at (capture.south) (captureusb) {\footnotesize USB};
  \node[rectangle, draw, align=center, minimum width=0.6cm, minimum height=0.5cm, anchor=west, color=gray] at ([yshift=+0.2cm]capture.west) (capturein) {\footnotesize in};

  \node[block, draw, align=center, minimum width=7.0cm, minimum height=1.2cm, text width=5.8cm, anchor=west] at (c) (pc) {~~\\~~\\Attack and measurement computer};
  \node[rectangle, draw, align=center, minimum width=0.8cm, minimum height=0.5cm, anchor=north, color=gray] at ([xshift=-2.4cm]pc.north) (pceth) {\footnotesize ETH};
  \node[rectangle, draw, align=center, minimum width=0.8cm, minimum height=0.5cm, anchor=north, color=gray] at ([xshift=2.4cm]pc.north) (pcusb) {\footnotesize USB};

  \draw [-{Stealth[scale=1.0]}, bend angle=20, bend left]  ([yshift=+0.2cm]dut.east) to node[above]{\footnotesize Electrical output} ([yshift=+0.2cm]capture.west);
  \draw [{Stealth[scale=1.0]}-, bend angle=20, bend right]  (duteth.south) to node[midway, right]{\footnotesize Attack traffic} (pceth.north);
  \draw [{Stealth[scale=1.0]}-{Stealth[scale=1.0]}, bend angle=20, bend left]  (captureusb.south) to node[midway, left]{\footnotesize Control \& measure} (pcusb.north);
  \end{tikzpicture}
  \caption{Test setup for the attack and measurement.}
  \label{fig:setup_overview}
\end{figure}
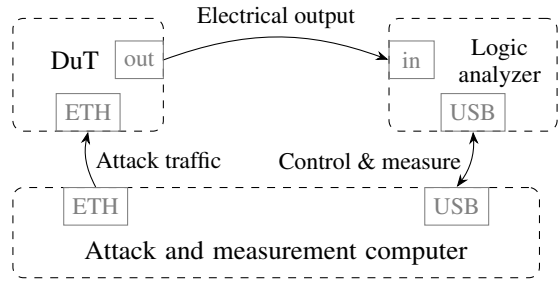

To measure the output signal of the \ac{DuT}, the Saleae Logic Pro logic analyzer is used \cite{saleae}, to capture the temporal progression of this signal during the test scenarios.
\Cref{fig:setup_overview} shows a schematic of the measurement setup.
The sample rate is configured to 250 Megasamples per second.
This device is connected to a computer, which is also executing the network stress tests via a direct Gigabit Ethernet connection to the \ac{DuT}.
Network hops in between might decrease the packet rate during a full load network attack, therefore we used a direct connection to simulate the worst case scenario, where the attacker has unlimited access without any bandwidth limitations.

\subsection{Attacking Tools}
\label{subsec:attackingtools}
Network flooding with SYN and \ac{ARP} packets and network scanning are carried out on the \ac{DuT} to disturb the regular execution.

For the first attack, a \ac{SYN} flood test generates a large amount of SYN packets, without handling the resulting responses from the \ac{DuT}.
The second attack is an \ac{ARP} flood program, which generates gratuitous \ac{ARP} requests, which are sent to the \ac{DuT}.

We also measure the impact of the common network scanning tool \texttt{nmap} \cite{nmap}.
The tool is used to detect configurations of the network connection, e.g. for asset management in the industrial network.
Without rate limitation this tool also generates massive network load for the \ac{DuT}.
In our test cases a full SYN scan for all 65535 ports is executed.
The use of this tool is not intended to disturb the receiver device, but since the scan of open ports without scan rate limitation can generate high network traffic, similar to the \ac{SYN} flood tool,
consequences for the cyclic control program under this network load were measured.
The measurement period is set to ten seconds for the idle and the attack measurements, due to the fact, that a full nmap scan on the \ac{DuT} takes around three to eight seconds in our setup.

Additional CPU load is generated to simulate high system utilization, e.g. brought by regular execution, demanding tasks or further attacks, which aim for high load generation on the device.

This system load is created with the common command line tool \texttt{stress-ng} and combined with the \ac{SYN} and \ac{ARP} flooding attacks and nmap network scanning.
Within this test run, we set one \ac{CPU} load process for single-core systems and two \ac{CPU} load processes for the dual-core systems, which get pinned to a \ac{CPU} each.
This results in a user-space process, which consumes 100 percent CPU time on all available cores.
Additional CPU load is intended to simulate high system utilization.
This is common for demanding industrial use cases, which require much processing time, at least for certain time.

\subsection{Measurement Procedure}
\label{subsec:measureprocedure}
A measurement procedure consists out of three measurements, with a duration of five minutes each.
While one measurement is taken during network attack, the other two capture the device in idle, before and after the attack.
A five seconds break is implemented between these three captures.
The measurement during the attack gives information about expectable consequences of the attack scenario, while the measurement before and after the attack is used for comparison to the regular execution and the idle jitter.
The capture after the attack also reveals information, if the influences of the attack are persistent, e.g. if the system crashed under the attack load.

\section{Results}
\label{sec:results}
In this section the measurement results of our setup are presented and outcomings are discussed.

\subsection{\acs{DoS} with Flooding for Single and Dual-core}
\begin{figure}[!ht]
   \begin{minipage}{0.49\columnwidth}
     \includegraphics[width=1.0\columnwidth]{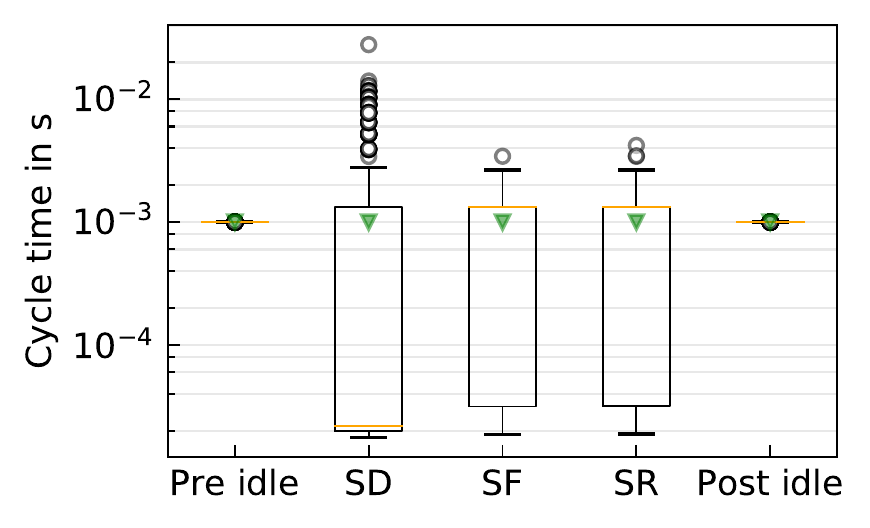}
     \caption{Single-core 5\,min SYN flood test for (SD), (SF) and (SR)}
     \label{fig:sc-5min-syn}
   \end{minipage}
   \hfill
   \begin{minipage}{0.49\columnwidth}
     \includegraphics[width=1.0\columnwidth]{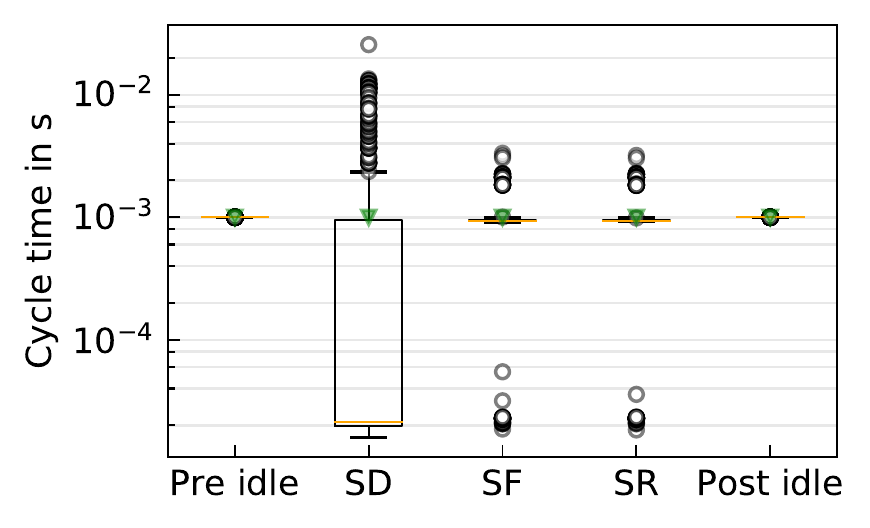}
     \caption{Single-core 5\,min ARP flood test for (SD), (SF) and (SR)}
     \label{fig:sc-5min-arp}
   \end{minipage}
 \end{figure}

The analysis with \ac{SYN} and \ac{ARP} based network attacks showed, that single-core configurations, like (SF), (SR) and (SD), have high outliers of their periodic toggle frequency during attack, in comparison to their idle measurement before and after.
The single-core setups do not show any deviation of mean cycle time, but have outliers multiple times higher and lower than the mean value.
There are outliers observed, which are multiple times the common cycle time of one millisecond, while the highest outliers can be found in the single-core setup with deadline scheduler (SD).
These high outliers get compensated by lower outliers, up to 62 times smaller, than the mean value.
This results in a mean cycle time without deviation, like the idle mean.
While faster cyclic execution should work well for most use cases, the higher outliers conditions delayed execution within the \ac{CPS}.
This behavior can cause severe disturbance to the controlled physical process.
Both network attacks result in comparable disturbance to the signal, especially for (SF) and (SR) the distribution differs and more cycles are found around the mean cycle time of 1\,ms.
The results for \ac{SYN} flooding attack on (SD), (SF) and (SR) are depicted in \Cref{fig:sc-5min-syn}.
Our measurement for \ac{ARP} flooding for the single-core test cases is shown in \Cref{fig:sc-5min-arp}.

\begin{figure}[!ht]
   \begin{minipage}{0.49\columnwidth}
     \includegraphics[width=1.0\columnwidth]{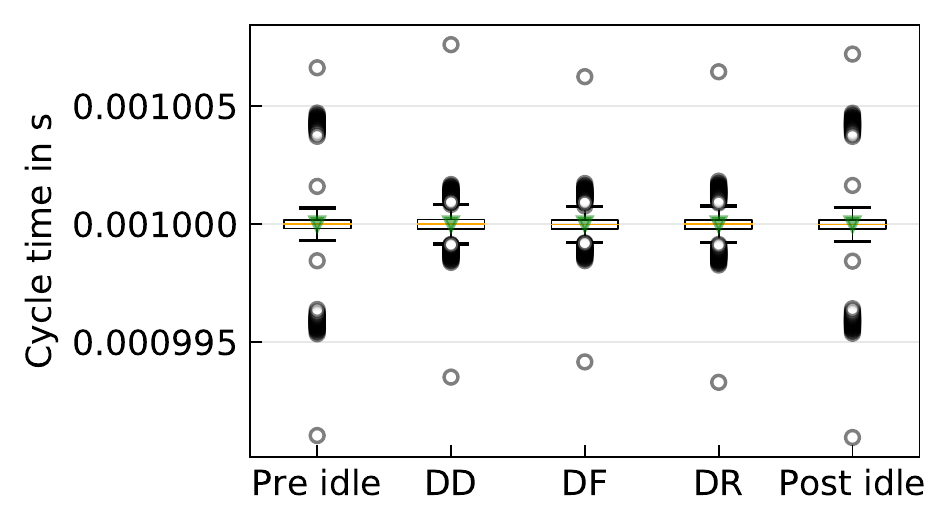}
   \caption{Dual-core 5\,min SYN flood test for (DD), (DF) and (DR)}
     \label{fig:dc-5min-syn}
   \end{minipage}
   \hfill
   \begin{minipage}{0.49\columnwidth}
     \includegraphics[width=1.0\columnwidth]{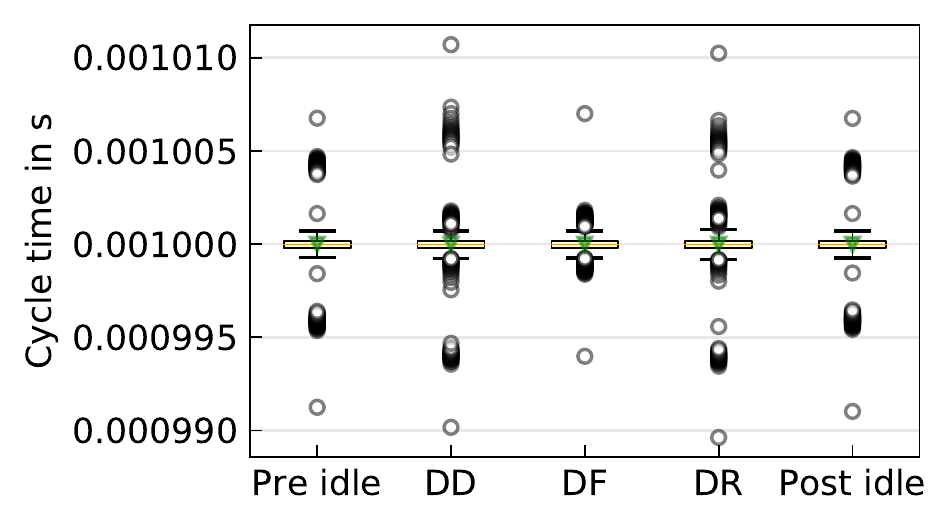}
   \caption{Dual-core 5\,min ARP flood test for (DD), (DF) and (DR)}
     \label{fig:dc-5min-arp}
   \end{minipage}
 \end{figure}

In the dual-core scenario with pinning the toggle process to CPU2, an impact on cycle time during the \ac{SYN} flood attack is not recognizable for all three test cases (DD), (DF) and (DR).
There are outliers measured during idle and attack, which jitter a few thousandths around the mean.
\Cref{fig:dc-5min-syn} shows the dual-core test cases during SYN flooding.
The dual-core test cases show minimal higher outliers during the \ac{ARP} flooding attack, but the deviation stays around one percent.
ARP flooding for all dual-core test cases is depicted in \Cref{fig:dc-5min-arp}.
Since the choice for a Linux-based system for \ac{RT} demands, presupposes the acceptance of some jitter during the execution of the control process, the measured dual-core systems provide a low jitter, even under network attack, for the configured cycle time.

\FloatBarrier
\subsection{\acs{DoS} with Flooding for Single and Dual-core and CPU Load}
In addition to network flooding, synthetic \ac{CPU} load is generated during the measurement on the \ac{DuT}, to simulate high system utilization.

\begin{figure}[!ht]
   \begin{minipage}{0.49\columnwidth}
     \includegraphics[width=1\columnwidth]{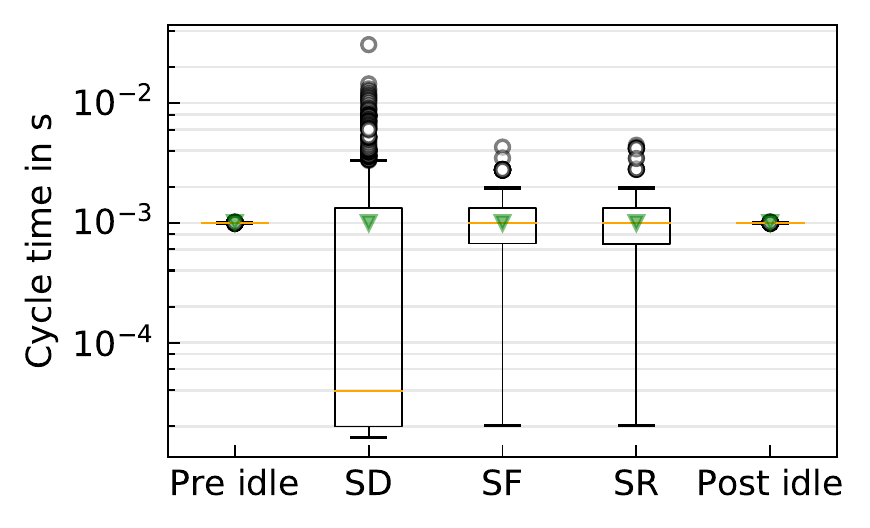}
     \caption{Single-core 5\,min SYN flood test for (SD), (SF) and (SR) with \ac{CPU} load}
     \label{fig:sc-5min-syn-stress}
   \end{minipage}
   \hfill
   \begin{minipage}{0.49\columnwidth}
     \includegraphics[width=1\columnwidth]{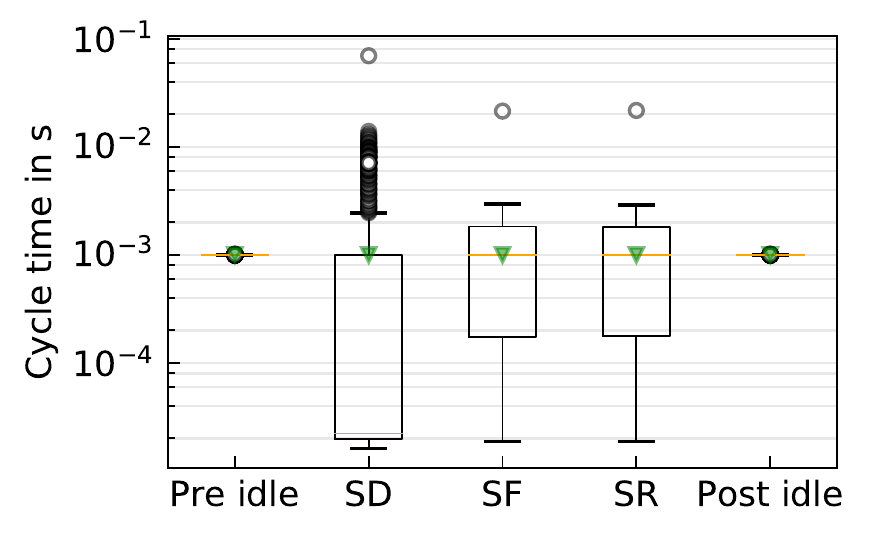}
     \caption{Single-core 5\,min ARP flood test for (SD), (SF) and (SR) with \ac{CPU} load}
     \label{fig:sc-5min-arp-stress}
   \end{minipage}
 \end{figure}

The single-core configurations (SD), (SF) and (SR) with stress show comparable distribution but higher outliers during ARP and SYN flooding, in comparison to the measurement without additional CPU load.
SYN flooding with stress is depicted in \Cref{fig:sc-5min-syn-stress}.
ARP flooding with stress is shown in \Cref{fig:sc-5min-arp-stress}.


The dual-core setups with additional stress now have higher outliers during \ac{SYN} and \ac{ARP} flooding.
This differs from the measurement of these systems without additional load, were only \ac{ARP} resulted minor outliers during attack.
The setups (DF) and (DR) now show higher and lower outliers during network attack, which results in a jitter around 11-13 percent.

(DD) with full CPU utilization, shows high outliers even in the idle measurements.
These outliers are two times the common cycle time and do not get worse under attack.
So a fully utilized dual-core system, using deadline scheduler shows very high jitter, even without additional network load.
A full utilized system using the deadline scheduler with our tested configuration does not provide a low jitter system, which seems usable for \ac{RT} controlling needs.
This odd behavior requires further investigation with this test case in addition to a fully loaded system.

\begin{figure}[!ht]
   \begin{minipage}{0.49\columnwidth}
     \includegraphics[width=1.0\columnwidth]{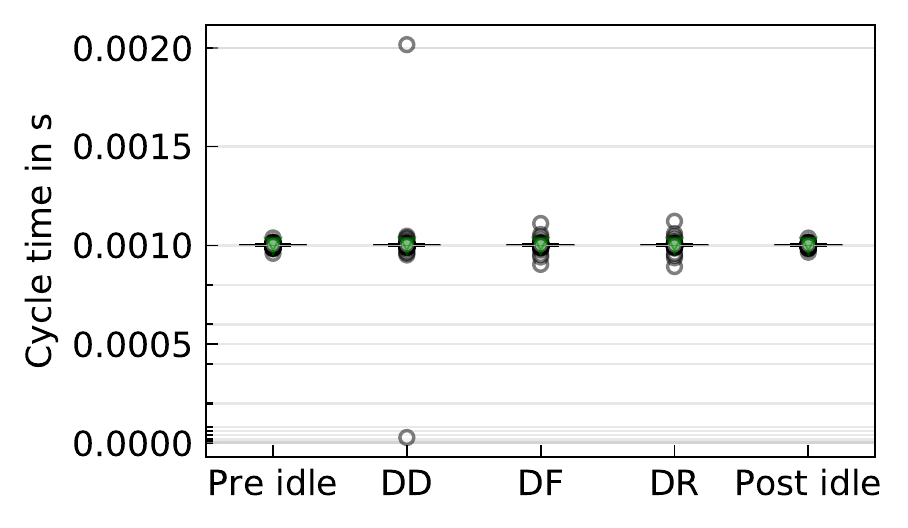}
     \caption{Dual-core 5\,min SYN flood test for (DD), (DF) and (DR) with \ac{CPU} load}
     \label{fig:dc-5min-syn-stress}
   \end{minipage}
   \hfill
   \begin{minipage}{0.49\columnwidth}
     \includegraphics[width=1.0\columnwidth]{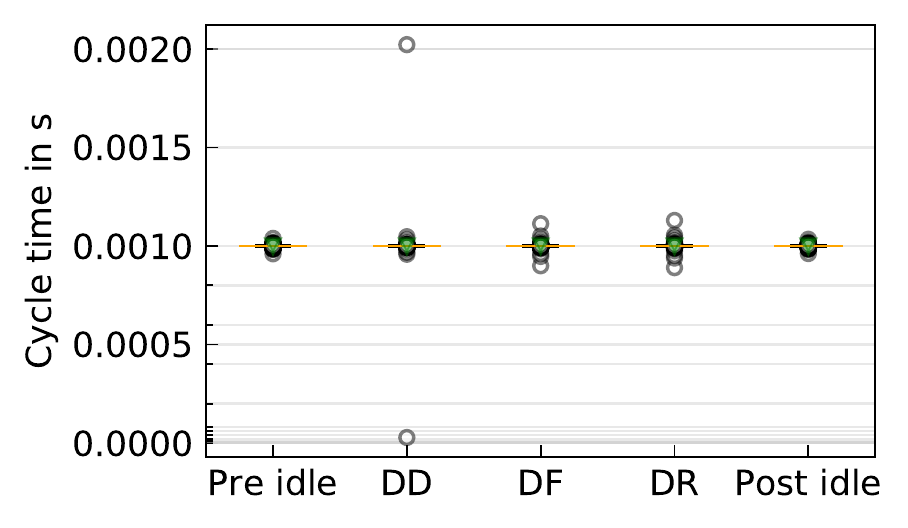}
     \caption{Dual-core 5\,min ARP flood test for (DD), (DF) and (DR) with \ac{CPU} load}
     \label{fig:dc-5min-arp-stress}
   \end{minipage}
 \end{figure}

SYN flooding on the dual-core test cases with additional \ac{CPU} load is depicted in \Cref{fig:dc-5min-syn-stress} and for ARP flooding in \Cref{fig:dc-5min-arp-stress}.
This shows also the odd behavior of (DD) under attack, which is similar in the pre and post idle.

\FloatBarrier
\subsection{Impacts of Network Scanning}
The previously shown scenarios are intended to provoke a \ac{DoS} of the \ac{DuT} intentionally.
However, network load is not just caused by offensive network traffic, but also by intentional network services, like scans.
Hence, this scenario shows the effects of a conventional network scanner (nmap) on the different architecture configurations.

\begin{figure}[htb!]
  \centering
  \includegraphics[width=0.6\columnwidth]{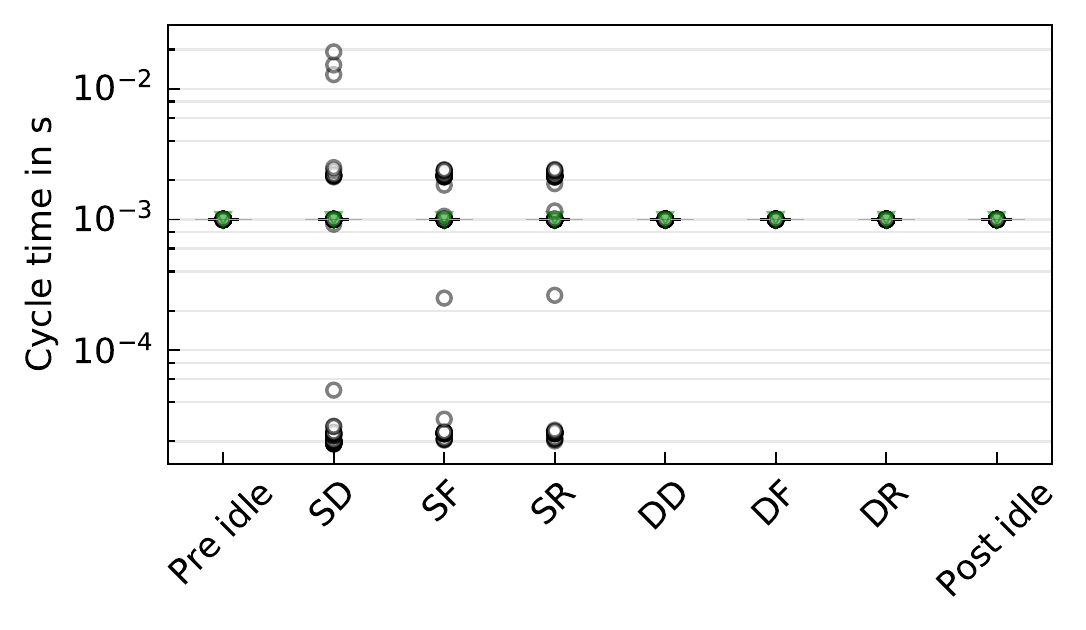}
  \caption{Nmap full scan 10\,s on single-core and dual-core test cases}
  \label{fig:nmap}
\end{figure}

The test cases (SD), (SF) and (SR) have similar outliers than under \ac{ARP} and \ac{SYN} flooding, while there is again no deviation from the mean cycle time.
Dual-core setups (DD), (DF) and (DR) do not show additional outliers during the nmap scan.
Resulting influences during nmap scanning is depicted in \Cref{fig:nmap}.

\subsection{Impacts on Co-Processor Architecture (C)}
\begin{figure}[htb!]
  \centering
  \includegraphics[width=0.6\columnwidth]{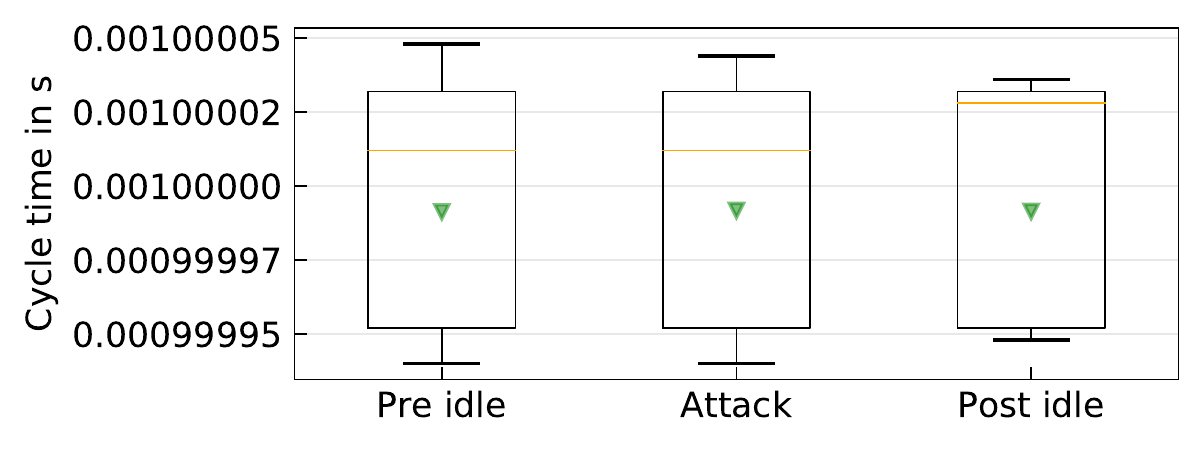}
\caption{Co-processor (C) 5min SYN Flood}
  \label{fig:copro-syn}
\end{figure}

The measurement of the co-processor test case showed no measurable deviation during all attacks, compared to idle.
All measured outliers in the measurements are within a very low jitter of a view nanoseconds.
This is conditioned by the strict separation between co-processor executing the cyclic program and the Linux system, which results in a near perfect cyclic signal.
\Cref{fig:copro-syn} depicts test case (C) under \ac{SYN} flooding attack, while the other attacks show similar results.

\FloatBarrier
\subsection{Network Packets and \acs{CPU} Load}

During the flooding attacks, the attacker sends about 100\,000 SYN packets/s to the \ac{DuT}.
The \ac{DuT} answers these packets with around 10\,000 SYN/ACK and SYN/RST packets/s.

This indicates, that not all network packets can be processed.
The flooding attack does not crash the network communication, since the \ac{DuT} constantly sends packets back to the sender device over the complete  measurement period.

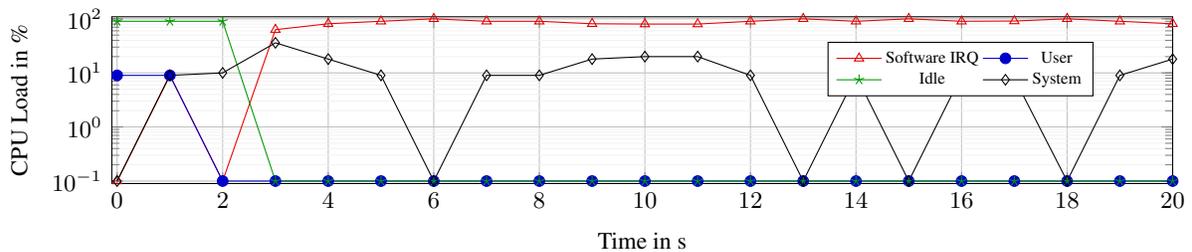
\begin{figure}[htb!]
	\centering
	\begin{tikzpicture}
	\tikzstyle{every node}=[font=\small]
	\begin{axis}[
	width=0.95\columnwidth,
	height=3.8cm,
	xlabel={Time in s},
	ymode=log,
	ylabel={CPU Load in \%},
	ymin = 0.09,
	xmin = -0.1,
	ymax = 109,
	xmax = 20,
	grid=both,
	grid style={line width=.1pt, draw=gray!10},
	major grid style={line width=.2pt,draw=gray!50},
	legend style={at={(0.8,0.85)},anchor=north, nodes={scale=0.7, transform shape}},
	legend columns=2,
	]
	\addplot+[red!90!black,mark=triangle,mark options={fill=red!90!black}] table [x=time,y={Software IRQ},col sep=comma] {./data/rpi_cpu.csv};
	\addlegendentry{\small Software IRQ};
	\addplot+[blue!80!black,mark=*,mark options={fill=blue!80!black}] table [x=time,y=User,col sep=comma] {./data/rpi_cpu.csv};
	\addlegendentry{\small User};
	\addplot+[green!60!black,mark=star,mark options={fill=green!60!black}] table [x=time,y=Idle,col sep=comma] {./data/rpi_cpu.csv};
	\addlegendentry{\small Idle};
	\addplot+[black,mark=diamond,mark options={fill=black}] table [x=time,y=System,col sep=comma] {./data/rpi_cpu.csv};
	\addlegendentry{\small System};

	\end{axis}
	\end{tikzpicture}
\caption{\acs{CPU} load during SYN flooding on the single-core Raspberry Pi 4.}
	\label{fig:CPUloadattack}
\end{figure}

To determine the cause of disturbance under network-based attacks, \ac{CPU} usage is analyzed for a single-core test case.
The \ac{CPU} load distribution during a SYN flooding attack is shown in \Cref{fig:CPUloadattack}.
It can be observed, that during the attack the software IRQ increases to almost 100\,\% CPU utilization.
This high utilization has an impact on the lower priority user tasks, e.g. the cyclic toggle program.

\section{Discussion}
\label{sec:discussion}

During our measurements the different architecture designs showed very different impacts to the different network-based and \ac{CPU}-based loads.
\Cref{tab:results} and \Cref{tab:resultsstress} show the highest measured outliers with and without additional \ac{CPU} load.
For future \ac{PLC} designs, the results can be a reference for the requirement of robustness.
The measured influences on the control process during network and system load determine the choice of the underlying architecture, therefore the measurement outcome gets summarized to give recommendations.

The introduced hardware separation with a co-processor provides best results in latency jitter, already in the idle state.
But the implementation of such an architecture requires low-level programming and software development skills.
To provide a deterministic and jitter free controller for \ac{CPS}, the co-processor solution is the only acceptable architecture from the measured test cases.
This makes the co-processor architecture the best choice for hard \ac{RT} demands, which can not accept single outliers during regular execution.
However, already dual-core systems, with additional task pinning, provide a robust setup for \ac{RT} demanding industrial control processes.
Even if there are influences measurable during attack or network scans, the measured dual-core setups provide overall a low jitter.
If little deviation from expected execution is tolerable for the control task, multi-core systems, running Linux \texttt{preempt\_rt} can offer a viable architecture for executing \ac{ICS} tasks.
Our measurements show, that task pinning in multi-core systems - (DD), (DF) and (DR) - provide robustness from network flooding attacks, since without additional load, no impact during attack can be measured.
Even in combination with \ac{CPU} load, (DF) and (DR) provide low cycle time delay, around 11-13 percent, while (DD) suffers from full utilization.
This architecture suits firm real-time needs, were some missed deadlines are tolerable and do not disturb regular execution.

\begin{table}[!htb]
  \centering
  \caption{Overview of maximum cycle time, without additional \ac{CPU} load}
  \small
  \label{tab:results}
    \begin{tabular}{l | l l l l}
    \hline
                & \textbf{Idle} & \textbf{\acs{SYN} flooding} & \textbf{\acs{ARP} flooding} & \textbf{Nmap}  \\
    \hline
    \textbf{SD}        & 1.008\,ms    & 27.852\,ms     & 25.685\,ms      & 21.678\,ms       \\ 
    \textbf{SF}        & 1.013\,ms    & 3.438\,ms      & 3.331\,ms       & 2.560\,ms       \\ 
    \textbf{SR}        & 1.008\,ms    & 4.213\,ms      & 3.200\,ms      & 2.603\,ms       \\ 
    \textbf{DD}        & 1.007\,ms    & 1.008\,ms      & 1.011\,ms       & 1.005\,ms       \\  
    \textbf{DF}        & 1.005\,ms    &  1.006\,ms     & 1.007\,ms       & 1.005\,ms       \\
    \textbf{DR}        & 1.006\,ms    &  1.006\,ms     & 1.010\,ms       & 1.005\,ms       \\
    \textbf{C}         & 1.000\,ms    &  1.000\,ms     &  1,000\,ms      & 1.000\,ms               \\
    \end{tabular}
\end{table}


\begin{table}[!htb]
  \centering
  \caption{Overview of maximum cycle time, with additional \ac{CPU} load}
  \label{tab:resultsstress}
  \small
    \begin{tabular}{l | l l l l}
    \hline
                & \textbf{Idle} & \textbf{\acs{SYN} flooding} & \textbf{\acs{ARP} flooding} & \textbf{Nmap}  \\
    \hline
    \textbf{SD} & 1.018\,ms    & 30.681\,ms     & 69.619\,ms     & 22.143\,ms       \\ 
    \textbf{SF} & 1.014\,ms    & 4.282\,ms      & 21.389\,ms     & 2.661\,ms       \\
    \textbf{SR} & 1.014\,ms    & 4.438\,ms      & 21.697\,ms     & 2.769\,ms       \\
    \textbf{DD} & 2.043\,ms    & 2.016 \,ms     & 2.022\,ms      & 2.027\,ms       \\ 
    \textbf{DF} & 1.039\,ms    & 1.111\,ms      & 1.113\,ms      & 1.070\,ms       \\
    \textbf{DR} & 1.052\,ms    & 1.122\,ms      & 1.130\,ms      & 1.082\,ms       \\
    \textbf{C}  & 1.000\,ms    &  1.000\,ms     &  1,000\,ms     & 1.000\,ms               \\
    \end{tabular}
\end{table}

Single-core solutions - (SD), (SF) and (SR) show high outliers during network-based attacks.
The cyclic execution shows outliers, which are multiple times higher, than the mean cycle time.
There are also outliers, which are multiple times smaller than this mean time.
While the higher outliers can cause massive delay in process execution, faster cycles are tolerable for most scenarios.
In addition, it should be mentioned, that the single-core test cases do not show any deviation in the mean value of the cycle time.
Therefore this architecture could be the correct choice for soft real-time control systems, if some high outliers of regular cyclic execution are neglectable.

\section{Conclusion}
\label{sec:conclusion}
In this work, we discussed how different configurations of schedulers and \ac{CPU} architectures influence the robustness of \ac{ICS} devices against high network communication loads.
For real-time control processes, a \texttt{preempt\_rt} patched Linux is a complex underlying system.
This results in a general higher jitter, due to kernel and scheduling latencies.
The use of multi-core systems with real-time patched Linux and pinning the critical process to a different CPU provides already good robustness for network flooding attacks, even in combination with high CPU utilization.

Physical processes, which are controlled by such devices, have specific but various demands on the \ac{RT} capabilities of the controlling device.
For these reasons, a general recommendation for a \ac{RT} capable industrial control device architecture is hard to define, since this choice also depends on factors, like development and hardware costs, required features of the underlying system and many more.

The outcome of this work can act as a reference to determine the choice of robust architectures for future \acp{PLC}.
Robustness of future \acp{PLC} against network-based attacks is essential, due to the increase of connectivity.
Therefore our measurement methodology and results should be considered for the selection of future architecture for \ac{RT} industrial devices.
If single deadline misses lead to catastrophic consequences for the control process, developers should reconsider the usage of Linux \texttt{preempt\_rt}, since our measurement shows high impact on single-core and measurable impact on multi-core test cases on the cycle time.
If this jitter is not tolerable for the control process or the controlled \ac{CPS} has hard \ac{RT} requirements, the execution of the control process on a dedicated \ac{CPU} is the only feasible solution within our measurements.

The measured heterogeneous multi-CPU architecture, provides Linux features and also a co-processor, which creates a predictable and robust \ac{RT} system.
But even multi-core Linux systems showed very little jitter, even under worst case network loads, which makes them suitable for multiple use cases.
If Linux-based systems are used for \ac{RT} process control, further additional mitigation strategies e.g. firewall rules must be considered and therefore can be part of future investigation and measurement.

\section*{Acknowledgments}
This work was partly funded by the Bavarian Ministry of Economic Affairs, Regional Development and Energy in the project ProLogCloud through grant number ESB066/003.

\bibliographystyle{unsrt}
\bibliography{bibliography.bib}  

%
%
%
%

\end{document}